\documentclass[aps,pre,twocolumn,superscriptaddress]{revtex4-2}
\usepackage{amssymb}
\usepackage{amsmath,bm}
\usepackage{graphicx,color}
\usepackage{bbm}
\usepackage[bottom]{footmisc}
\usepackage{multirow}
\usepackage{xcolor}
\usepackage{xpatch} 
\makeatletter   
\usepackage{xpatch} 

\newcommand{\B}[1]{{\bm{#1}}}

\newcommand{\Lag}{\mathcal{L}}
\newcommand{\A}{\mathcal{A}}
\newcommand{\dif}{\mathrm{d}}
\newcommand{\rin}{r_\text{in}}
\newcommand{\rout}{r_\text{out}}
\newcommand{\C}[1]{{\mathcal{#1}}}

\begin{document}
\title{Anomalous Elasticity and Screening in Amorphous Solids}
\author{Ana\"el Lema\^{\i}tre}
\affiliation{NAVIER, UMR 8205, \'Ecole des Ponts ParisTech, IFSTTAR, CNRS, UPE, Champs-sur-Marne, France}
\author{Chandana Mondal}
\affiliation{Dept. of Chemical Physics, The Weizmann Institute of Science, Rehovot 76100, Israel}
\author{Michael Moshe$^\dag$}
\email{michael.moshe@mail.huji.ac.il}
\affiliation{Racah Institute of Physics, The Hebrew University of Jerusalem, Jerusalem, Israel 9190}
\author{Itamar Procaccia$^\dag$} 
\affiliation{Dept. of Chemical Physics, The Weizmann Institute of Science, Rehovot 76100, Israel, $^*$Center for OPTical IMagery Analysis and Learning, Northwestern Polytechnical University, Xi'an, 710072 China}
\author{Saikat Roy} 
\affiliation{Department of Chemical Engineering, Indian Institute of Technology Ropar, Punjab 140001, India}
\author{Keren Screiber-Re'em}
\affiliation{Racah Institute of Physics, The Hebrew University of Jerusalem, Jerusalem, Israel 9190}

\begin{abstract}
Amorphous solids appear to react elastically to small external strains, but in contrast to ideal elastic media, plastic responses abound immediately, at any value of the strain. Such plastic responses are quasi-localized in nature, with the ``cheapest" one being a quadrupolar source. The existence of such plastic responses results in {\em screened elasticity} in which strains and stresses can either quantitatively or qualitatively differ from the un-screened theory, depending on the specific screening mechanism. Here we offer a theory of such screening effects by plastic quadrupoles, dipoles and monopoles, explain their natural appearance, and point out the analogy to electrostatic screening by electric charges and dipoles.  For low density of quadrupoles the effect is to normalize the elastic moduli without a qualitative change compared to pure elasticity theory; for higher density of quadrupoles the screening effects result in qualitative changes. Predictions for the spatial dependence of displacement fields caused by local sources of strains are provided and compared to numerical simulations. We find that anomalous elasticity is richer than
electrostatics in having a screening mode that does not appear in the electrostatic analog.\\

$^\dag$ Corresponding author

\end{abstract}
\maketitle
\section{Introduction}

Solid matter differs from liquids in its ability to support stress, and the linear theory of elasticity is the framework that attempts to quantify the response of matter to small deformations \cite{86LKP}.  In isotropic and homogeneous solids, the shear and bulk moduli, for example, relate the stress response to the external strain. In perfect elastic materials one can also consider nonlinear responses, quantified in terms of nonlinear moduli which relate large deformations to higher order stress responses.

In amorphous solids the situation is less straightforward. Experimentally one observes that stresses increase linearly with strains at small deformations, providing an impression that such solids comply with
linear and nonlinear elasticity theory as well as their crystalline counterparts. In fact, research in the last few decades indicates that this is not the case. Firstly, it was shown that amorphous solids suffer from plastic responses \cite{99ML,06ML}, and in the thermodynamic limit these appear for any infinitesimal deformation \cite{10KLPb,11HKLP}. Secondly, the nature of these plastic responses, being quasi-localized, results in the destruction of nonlinear elasticity; the higher order elastic moduli have sample-to-sample fluctuations which are not converging with the increase of the system size; quite on the contrary, these can even diverge \cite{11HKLP,16PRS,16DPSS,17DIPS}.
Under external shear strain the accumulation of plastic responses can lead to mechanical failure of amorphous solids through shear-banding and the appearance of cracks \cite{12DHP,13DHP}. 

Having these anomalies in mind, the aim of this paper is to provide a theory of what we term ``anomalous elasticity", which in the context of amorphous solids takes into explicit account the existence of prevalent plastic responses. In particular we will argue that the existence of such responses lead to interesting screening phenomena that appear not to have been studied in the context of amorphous solids mechanics. The concept of screening is common to any student of electrostatics \cite{84LL}. Indeed, the well known Debye-H\"uckel approach provided a theoretical explanation for departures from ideality in solutions of electrolytes and plasmas. This theory considered the existence of mobile charges (or monopoles) in solutions, leading to screening effects resulting from the electrostatic interactions between ions and their surrounding clouds. An equally important subject is dielectrics, in which an external electric field induces
electric dipoles which in turn interact to screen the inducing field. In the case of electrostatics further multipole expansion were deemed unnecessary in normal statistical physics. We note that these two screening mechanisms are different, the first caused by existing charges and results with qualitatively stronger
screening effects compared to the dielectric example, in which only renormalization of the dielectric
constants is being observed \cite{84LL}. One way to effectively distinguish between the types of screening is to
measure the spatial decay of the responding field to a local charge. In this paper we examine an analogous approach, considering the responding displacement field
of an amorphous solids to a local elastic charge. 

We argue here that in analogy to electrostatics, the screening charges are induced and are plastic in nature.  We will study systems in mechanical equilibria in which a source of strain is added, and the main question will be what is the resulting screened displacement field. As is well known, in normal elasticity induced displacement field decay at long distances like $1/r^{D-1}$ where $r$ is the distance from the source and $D$ is the space dimension \cite{21LMPR}. We will see that in amorphous solids the functional shape of the displacement field depends intimately on the {\em density} of quadrupolar plastic responses. When their density is low, only renormalization of the elastic moduli is observed, in analogy to dielectrics  \cite{20NWRZBC}. Increasing their density, quadrupoles cooperate to form effective dipoles, where qualitatively new screening effects will be predicted and measured, with no immediate electrostatic analog. Finally at extremely high densities of plastic
deformation monopoles form, and similar screening effects to Debye-H\"uckel are expected  to exist. In fact, such a situation is tantamount to melting the amorphous solids, bringing us beyond strict anomalous elasticity and therefore beyond the scope of the present paper. This sequence of screening phenomena is analogous to the melting of two-dimensional 
crystals. At low temperatures only dipole pairs (quadrupoles) renormalize the elastic moduli,
then at a critical temperature quadrupoles unbind to form the hexatic phase, in which dipoles
are the screening objects \cite{80ZHN}. Finally at the melting temperatures monopoles unbind from the dipoles and
the solid structure collapses. The difference from the crystalline analog is that the screening objects are not structural, but mechanical, allowing us to formulate an intermediate "hexatic" phase in an isotropic and homogeneous medium.

\begin{figure}
		\includegraphics[width=0.7\linewidth]{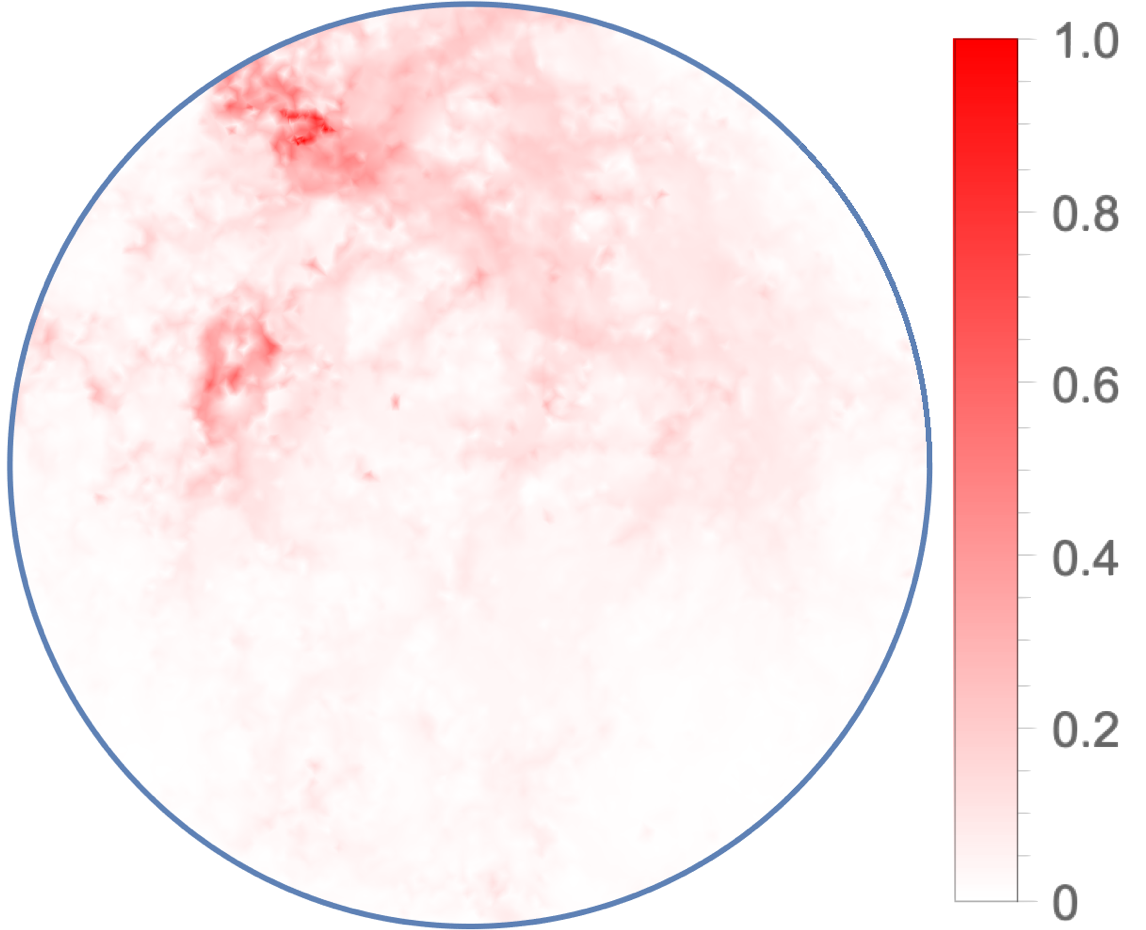}
		\includegraphics[width=0.7\linewidth]{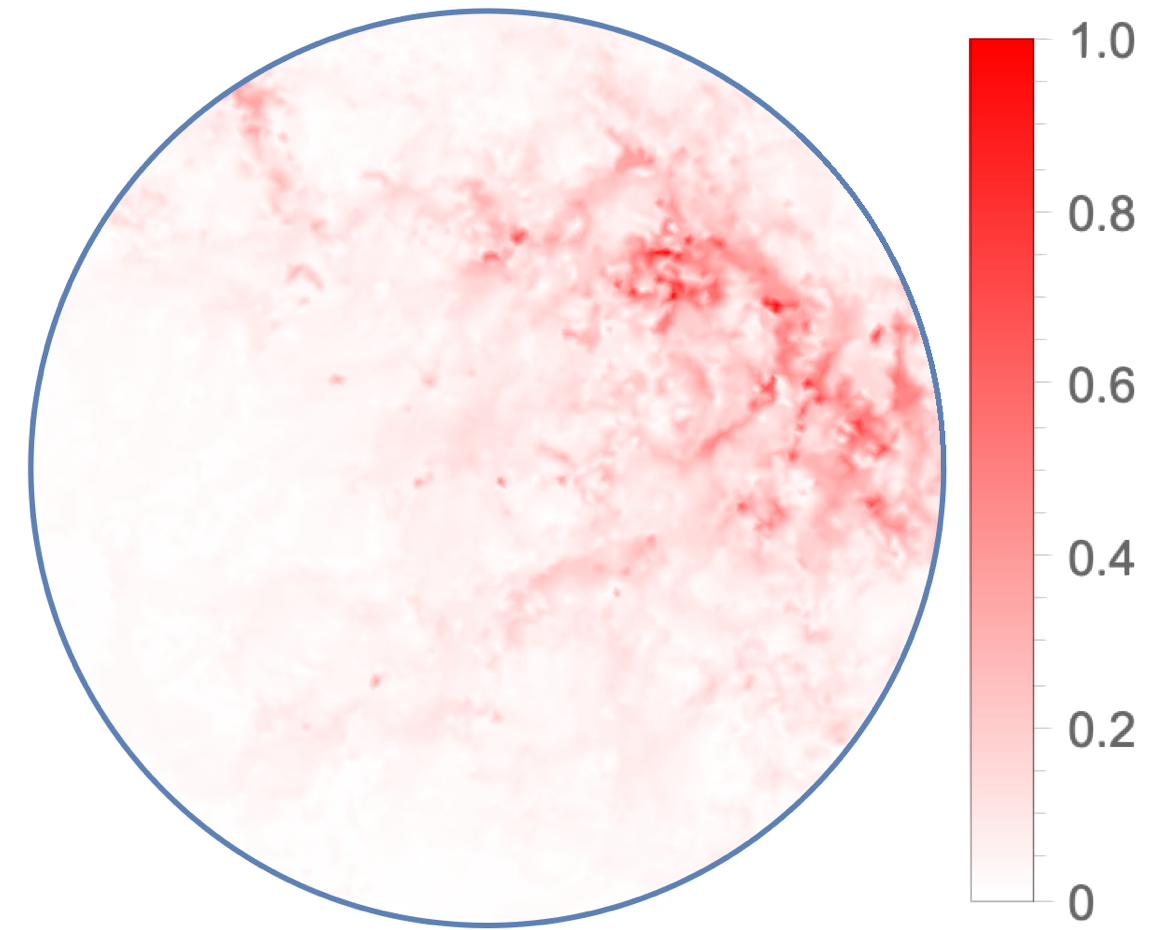}
	\caption{Displacement field induced by the inflation of one disk at the center of the box by 1\%. The displacement  vectors are normalized by the maximal one and plotted with the color code shown on the right. These are two different configurations prepared with identical protocols, and see Appendix~\ref{proto} for details.}
	\label{disP4.5}
\end{figure}

\begin{figure}
	\includegraphics[width=1.1\linewidth]{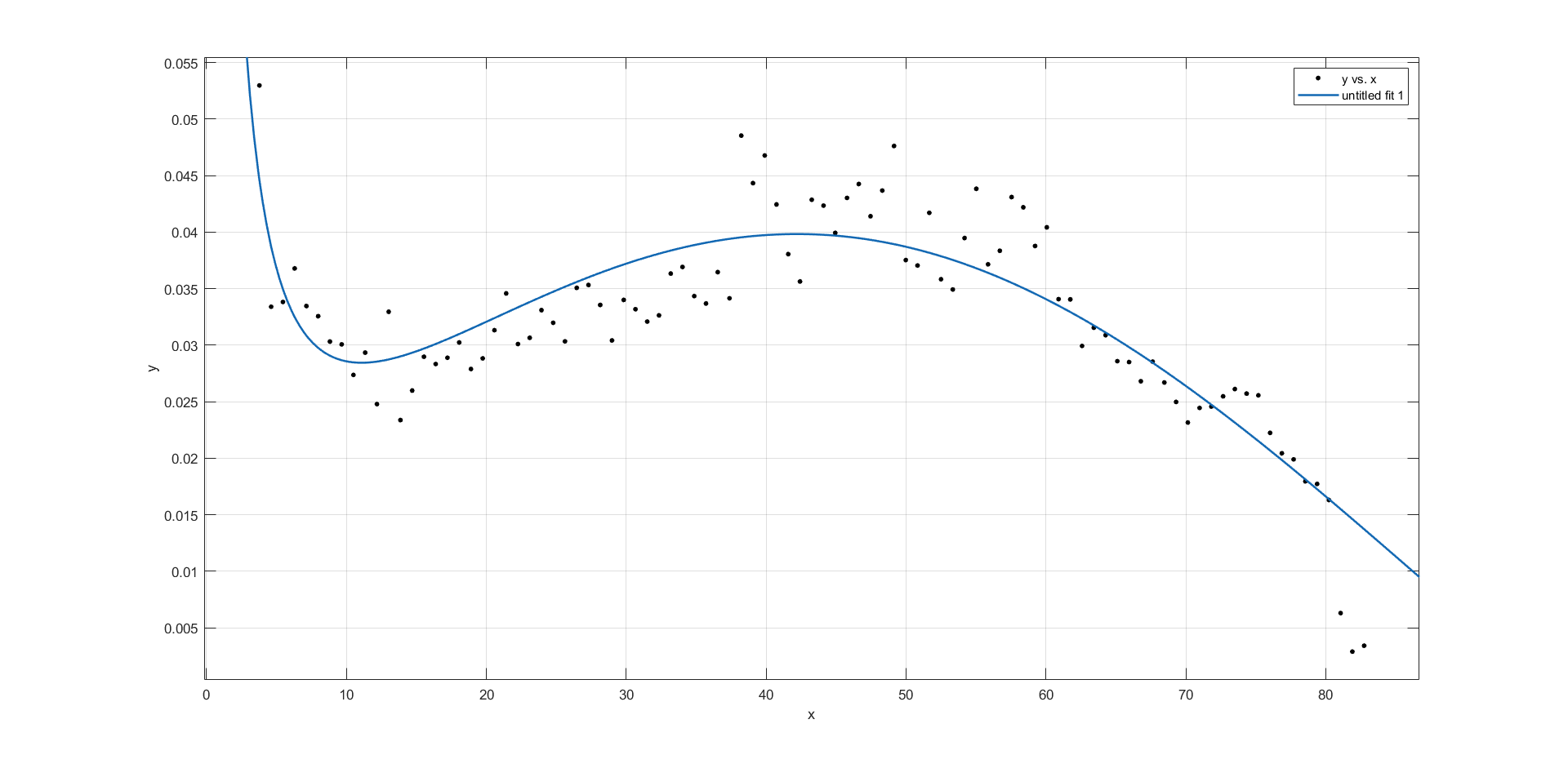}
	\includegraphics[width=1.1\linewidth]{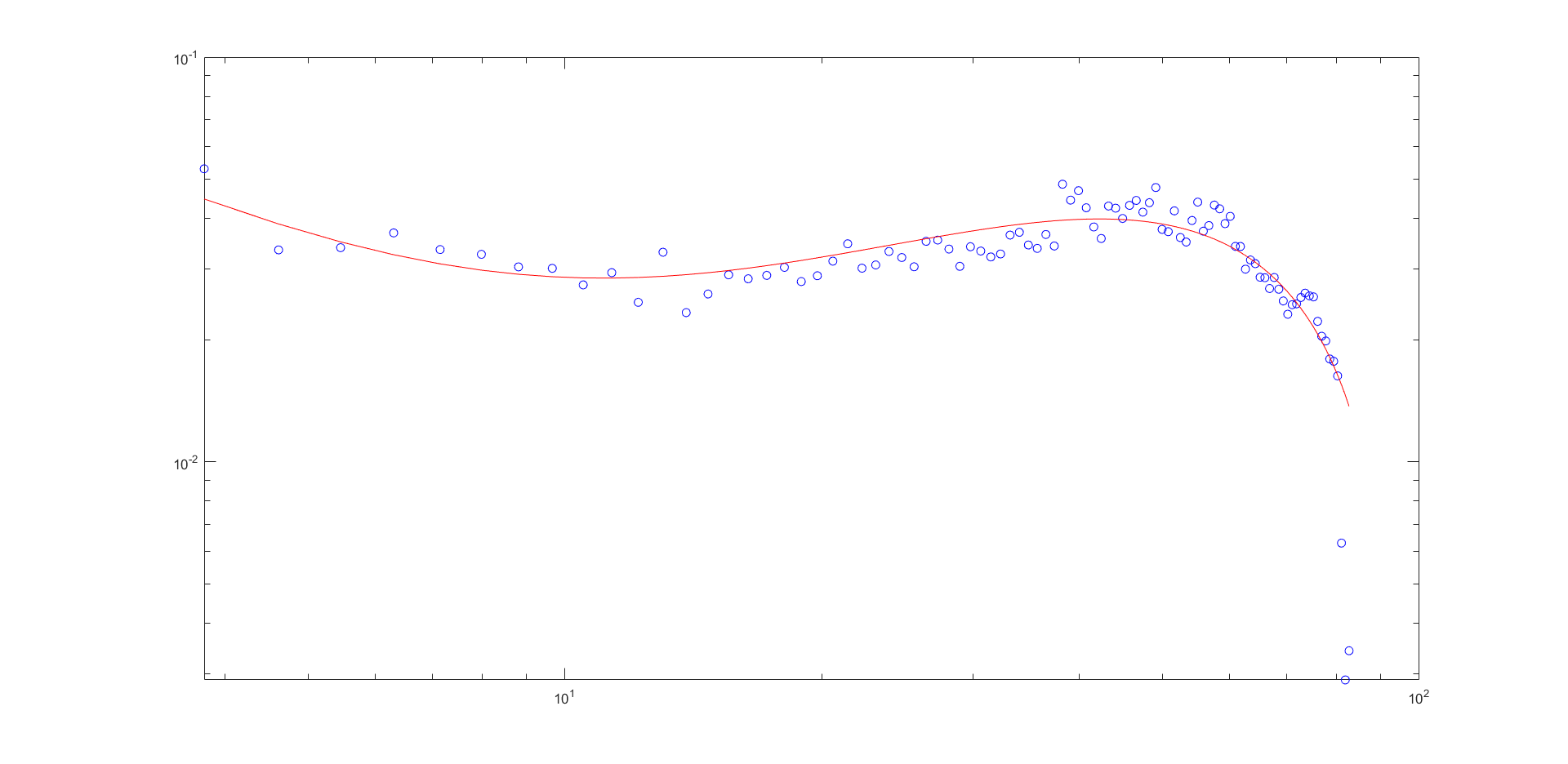}
	\caption{Angle averaged displacement field computed from the data
	presented in Fig.~\ref{disP4.5}. Together with the data we present in continuous lines the theoretical predictions that are developed for the angle averaged displacement in this paper. The parameters used in Eq.(\ref{amazing}) to fit the data are $\rin=3.9391,k= 0.0381,d_0=0.0326,\rout=98.3690$ for the upper panel and $\rin= 4.9819,k= 0.0380,d_0= 0.0368, \rout= 97.1784$ in the lower panel.}
	\label{angav}
\end{figure}

To illustrate the emerging theory we will consider the response to a local stress increase in an amorphous solid made from a binary mixture of disks
which are contained in a circular box, and see Appendix~\ref{proto} for details of interactions and protocol of preparation. Once brought to mechanical equilibrium at a target pressure, a disk is chosen closest to the center of the box. We then inflate this disk by a small percentage of 1\% , and examine the displacement field that is induced by this inflation. In a normal elastic material in radial geometry we expect the displacement field to decay like $1/r$. Typical displacement fields obtained form pressure $P=4.5$ (for units see Appendix~\ref{proto}) and inflation of $1\%$ are shown in Fig.~\ref{disP4.5}.
Clearly, the displacement field does {\em not} appear to decay as expected in normal elastic materials. Examining the radial dependence of the angle-averaged displacement, we find the results reported in Fig.~\ref{angav}.
Indeed, we see that the displacement field can even {\em increase} when
$r$ increases, in a striking contradiction with the normal elasticity
expectation. We will see below that the theory developed in this paper
explains fully and quantitatively this observed behavior and other
similar novel results. Indeed, the continuous lines in Fig.~\ref{angav} represent the analytic theory that is developed in the present paper. 

The structure of this paper is as follows: In Sect.~\ref{elastic} we remind the reader the standard theory of the spatial dependence of displacement fields induced by a local elastic charge.  The third Sect.~\ref{dilquad} will discusses screening by a dilute set of quadrupolar responses. We will see that dilute plastic responses can only renormalize the elastic moduli. Sect.~\ref{robust}
deals with the situation of higher densities of quadrupoles that aggregate to effective dipoles. This section will lead to the theory of anomalous
elasticity that can explain the results shown in Figs.~\ref{disP4.5} and \ref{angav}.
The effects of monopoles are of less immediate interest for us, being related to melting. In Sect.~\ref{PorE} we demonstrate that the response that results in the observed screening is indeed plastic and not non-affine elastic.  The last section \ref{summary} will offer a summary of the paper and indications for the road ahead. 

\section{Displacement fields induced by local isotropic charges in purely elastic matter}
\label{elastic}

The aim of this Section is to remind the reader of the normal expected response of an elastic system to a local charges. We start with the minimization of the elastic energy. 

\subsection{Minimizing the total energy}

The general strategy throughout the rest of this paper is to obtain the equation satisfied
by the displacement field by minimizing the total energy of the system \cite{13DHP}. When the system is purely elastic the energy $F$ can be represented as a sum of two contributions, an internal energy $U$ and the work done by traction forces $\B t$ on the boundary, denoted $W$. $U$ is an integral over the stress times the strain. The strain and stress tensors are denoted $u_{\alpha\beta}$,  $\sigma^{\alpha\beta}$ and the displacement field by $d_\alpha$. The energy density will be denoted as $\C L$ and in two dimensions the total energy $F$ is 
\begin{eqnarray}
F &=& \int \Lag \, \mathrm{d}^2 x - \oint t^\beta  d_\beta \, \mathrm{d}l \ , \nonumber\\
\C L &=&  \frac{1}{2} A^{\alpha\beta\gamma\delta} u_{\alpha\beta}u_{\gamma\delta} =\frac{1}{2}\sigma^{\alpha\beta} u_{\alpha\beta} \ .
\label{basic}
\end{eqnarray} 
The strain is related with the displacement field via 
\begin{equation}
 u_{\alpha\beta} = \frac{1}{2}\left(\partial_\alpha  d_\beta  + \partial_\beta d_\alpha \right) \ .
\end{equation} 

Next we minimize the total energy with respect to $\B d$ 
\begin{equation}
\begin{split}
\delta_d F&= \delta_d \int \Lag \dif^2 x - \oint t^\beta \delta d_\beta \mathrm{d}l \\&=\int \dif^2 x  \sigma^{\alpha\beta}\delta u_{\alpha\beta}- \oint t^\beta \delta d_\beta \mathrm{d} l \ .
\end{split}
\label{StrainVar1}
\end{equation} 
We note that
\begin{equation}
\delta u_{\alpha\beta} = \frac{1}{2}\left(\partial_\alpha \delta d_\beta  + \partial_\beta \delta
 d_\alpha \right) \ .
 \label{strainvar}
\end{equation} 
Substituting in Eq.(\ref{StrainVar1}) and integrating by parts we get
\begin{equation}
\begin{split}
\delta_d F &= \int \dif^2 x \sigma^{\alpha\beta} \partial_\alpha \delta d_\beta- \oint t^\beta \delta d_\beta \mathrm{d} l\\ &= \oint \left(\sigma^{\alpha\beta} \, n_\alpha  - t^\beta\right) d_\beta \dif \ell  -  \int \dif^2 x \partial_\alpha \sigma^{\alpha\beta} \delta d_\beta 
\end{split}
\end{equation} 
hence
\begin{eqnarray}
\partial_\alpha \sigma^{\alpha\beta} &=& 0\nonumber\\
\sigma^{\alpha\beta} \, n_\beta \lvert_{\partial} &=& t^{\alpha} \ .
\label{equi}
\end{eqnarray}
Upon substituting the expression for the stress tensor in terms of strain, and then using the relation of strain and displacement, we find the equation for the displacement field in the form 
\begin{equation}
\Delta \mathbf{d} + \lambda \nabla  \left(\nabla \cdot \mathbf{d}\right)= 0 \ , \quad \lambda\equiv \frac{1+\nu}{1-\nu} \ ,
\label{eq:EquilibriumDisp}
\end{equation}
where $\nu$ is the 2-dimensional Poisson ratio. 

\subsection{Example: Isotropic inclusion}
\begin{figure}
	\includegraphics[width=0.9\linewidth]{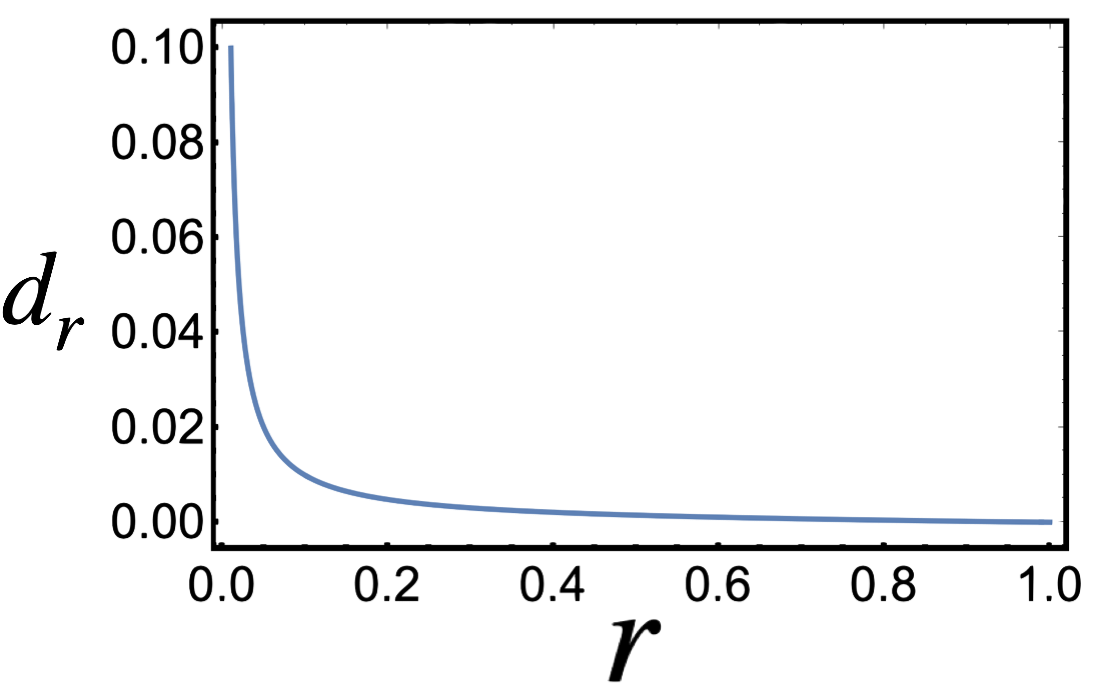}
	\caption{Graphic representation of the solution of Eq.~(\ref{renelas}) for $\rin=1$ and $\rout=10$.}
	\label{simple}
\end{figure}

In preparation of the more interesting solutions in the sections ahead, we consider here an annulus of radii $\rin$ and $\rout$ with an imposed displacement $\mathbf{d}(\rin) = d_0 \hat{r}$ and $\mathbf{d}(\rout) = 0$. The polar symmetry of the problem implies that $\mathbf{d}(r) =d_r(r) \hat{r}$, in which case the equilibrium equation reduces to 
\begin{equation}
	\Delta {\B d}=0 \ .
\end{equation}
The solution to this differential equation is
\begin{equation}
	{d}_r(r)=d_0 \frac{r^2 - \rout^2}{\rin^2 - \rout^2}\frac{\rin}{r} \ .
	\label{renelas}
\end{equation}
The graphic representation of this solution is exhibited in Fig.~\ref{simple}, 
showing that the solution goes like $1/r$ as is expected in standard elasticity theory.

\section{The effect of dilute quadrupoles}
\label{dilquad}

In this section we consider the effect of dilute quadrupoles. The final 
result will be elasticity theory with renormalized elastic moduli. This is
already an emergent theory, since the moduli are determined self-consistently by the response of the system. 

\subsection{Background: Elastic potential}

Eq.~(\ref{equi}) means that the stress field is divergence free.  
A divergence free field can always be represented in terms of an elastic (Airy's) potential \cite{73Gur}:
\begin{eqnarray}
	\sigma^{\alpha\beta}  = \varepsilon^{\alpha\mu} \varepsilon^{\beta\nu} \partial_{\mu\nu} \chi \ ,
\end{eqnarray} 
with $\B \varepsilon$ the anti-symmetric tensor.
The compatibility condition on the strain tensor, which reflects the inter-dependence of its components, is
 \begin{eqnarray}
 	\varepsilon^{\alpha\mu} \varepsilon^{\beta\nu} \partial_{\mu\nu} u_{\alpha\beta} = 0 \ .
 \end{eqnarray} 
Upon combining the stress representation with the compatibility condition we find
 \begin{eqnarray}
	\Delta\Delta \chi = 0 \ .
\end{eqnarray} 
This is the well known bi-harmonic equation that was treated extensively in the literature, cf. \cite{53Mus}.

\subsection{Background: Elastic charges}
Sources of stress can be incorporated into the theory by adding them to the bi-harmonic equation
  \begin{eqnarray}
 	\Delta\Delta \chi(x) = \rho(x)
 \end{eqnarray} 
Generally speaking, the source function $\rho(x)$ depends on how the source is introduced. But in the far field approximation we can consider the sources as singular for the bi-harmonic equation. Singular sources can be represented by a multipole expansion according to 
  \begin{eqnarray}
	\rho(x) = m \, \delta(x) + b^\alpha \partial_\alpha \delta(x) +  q^{\alpha\beta} \partial_{\alpha\beta} \delta(x) + \dots \ .
\end{eqnarray} 
The monopole and dipole charges $m$ and $\B b$ describe disclinations and dislocations, which are topologically protected and cannot form locally. The quadrupole term describes the first non-topological source with the charge $\B q$ related to the well known eigen-strain of an Eshelby inclusion \cite{54Esh}. The quadrupolar charge has the dimension of an area. Below we will denote the magnitude of the eigen-strain as $q$.
An extensive presentation and discussion of elastic charges within the linear theory is given in Ref.~\cite{15MSK,21SCBK,21SCBKa}. A generalization to nonlinear elasticity can be found in Ref.~\cite{20BLBM}.

Of particular importance is the Green's function corresponding to the quadrupolar term, solving the equation
  \begin{eqnarray}
	\Delta\Delta G_{\alpha\beta}(x) = \partial_{\alpha\beta} \delta(x) \ .
\end{eqnarray} 
Using the linearity of the equation we relate $G_{\alpha\beta}(x)$ with the derivatives of the monopole Green's function \cite{15MSK}
\begin{eqnarray}
G_{\alpha\beta}(x) = \frac{1}{16\pi}   \partial_{\alpha\beta} \left[  \lvert x  \rvert^2 \left( \log \lvert x  \rvert^2 - 1 \right)\right] \ .
\end{eqnarray}

\subsection{Microscopic description and mean-field theory}
\label{micro}
When the density of quadrupoles is low, there exist a scale separation $\sqrt{q} \ll \ell_q \ll L$. Here $\ell_q$ is the typical distance between quadrupoles, and $L$ is the system size.
The total quadrupole charge in a small area $\Delta S$ is
\begin{eqnarray}
	 q_\text{tot}^{\alpha\beta} = \sum_{x_i \in \Delta S} q^{\alpha\beta}_i \delta (x - x_i)
\end{eqnarray}
The elastic potential induced by a quadrupolar charge $q^{\alpha\beta}$ located at $x'$ is denoted $G_{\alpha\beta} (x - x')$, hence the total elastic potential is
\begin{equation}
		\chi = \sum_i q_i^{\alpha\beta} G_{\alpha\beta} (x - x_i)\ .
		\label{defchi}
\end{equation}
In the case of low quadrupole density Eq.~(\ref{defchi}) is re-written in the continuum limit in the form

\begin{equation}
		\chi(x)  =   \int \dif ^2 x' \, Q^{\alpha\beta}(x') G_{\alpha\beta} (x-x') \ ,
	\label{chicont}
\end{equation}
where $ \dif ^2 x' \, Q^{\alpha\beta}(x') \equiv q_{\rm tot}^{\alpha\beta}$

We define the stress and strain Green's functions 
\begin{eqnarray}
	G^{\mu\nu}_{\quad\alpha\beta}(\sigma;x - x') = \varepsilon^{\mu\rho}\varepsilon^{\nu\eta} \partial_{\rho\eta} G_{\alpha\beta}(x - x')
	\nonumber\\
	G_{\mu\nu\alpha\beta}(u;x - x')  = \A_{\mu\nu\rho\eta} G^{\rho\eta}_{\quad\alpha\beta}(x - x') \ .
\end{eqnarray} 
From these Green's function we can compute the extra strain and stress which are induced by the
quadrupole field, denoted as $\sigma^{\mu\nu}(x|Q)$, $u_{\mu\nu}(x|Q)$:
\begin{eqnarray}
	\sigma^{\mu\nu}(x|Q)&=  \int \dif^2 x'  G^{\mu\nu}_{\alpha\beta}(\sigma;x - x') Q^{\alpha\beta} (x') 
	\nonumber\\
u_{\mu\nu}(x|Q)&=  
	\int \dif^2 x'  G_{\mu\nu\alpha\beta}(u;x - x') Q^{\alpha\beta} (x') \ .
\end{eqnarray}

\subsection{Mean field equations}
In general, the total strain in a given system is not determined only by the density of quadrupoles, there can be additional mechanical sources of strain. We therefore write the total strain in the system as:
\begin{equation}
	u_{\alpha\beta} (x)= u^\text{el}_{\alpha\beta}(x) + u_{\alpha\beta} (x|Q) \ ,
	\label{eq:straindecomp}
\end{equation} 
where $u^\text{el}_{\alpha\beta}(x)$ is the ``normal" strain field. 
The energy functional, from which the equilibrium equations are derived, contains now contributions from both the quadrupoles and the normal elastic fields  \cite{13DHP}. In other words, 
\begin{equation}
\begin{split}
U = U_\text{el} + U_\text{QQ} + U_\text{Q-el} \ .
\end{split}
\label{eq:Udecomp}
\end{equation} 
Finding the actual form of these three interactions requires starting from the detailed energetic considerations as done for example in Ref.~\cite{13DHP}. This calculation is presented in Appendix~\ref{micro}, with the final result 
reading
\begin{equation}
\begin{split}
	U_\text{el} &=\int \dif^2 x \frac{1}{2} A^{\alpha\beta\gamma\delta} u_{\alpha\beta}u_{\gamma\delta}\\
	U_\text{QQ} &=\int \dif^2 x \frac{1}{2} \Lambda_{\alpha\beta\gamma\delta} Q^{\alpha\beta}Q^{\gamma\delta}\\
	U_\text{Q-el} &=  \int \dif^2 x \Gamma^{\alpha\beta}_{\gamma\delta} u_{\alpha\beta}Q^{\gamma\delta} \ .
\end{split}
\label{eq:energydecomp}
\end{equation} 
We note that in principle neither $u_{\alpha\beta}$ nor $Q^{\alpha\beta}$ are a-priori known even when the external loads are given. Both will have to be computed by minimizing the energy with respect to them.

\subsection{Minimization of the energy functional}
 Upon minimizing \eqref{eq:Udecomp} with respect to the fundamental fields $d$ and $Q$ we find 
\begin{equation}
	\begin{split}
		\delta_Q U= \delta_Q \int \Lag \dif^2 x=  \int \dif^2 x \left( \Lambda_{\alpha\beta\gamma\delta} Q^{\alpha\beta}+ \Gamma^{\alpha\beta}_{\gamma\delta} u_{\alpha\beta}\right)\delta Q^{\gamma\delta}\\
		\delta_d U= \delta_d \int \Lag \dif^2 x=  \int \dif^2 x \left( \sigma^{\alpha\beta}\delta u_{\alpha\beta} +   \Gamma^{\alpha\beta}_{\gamma\delta} Q^{\gamma\delta} \delta u_{\alpha\beta}\right)  
	\end{split}
	\label{StrainVar}
\end{equation} 
From the first equation we get a linear screening relation (analogous to the linear relation between electric field and induced polarization in dielectric materials)
\begin{equation}
	Q^{\alpha\beta} = - \Lambda^{\alpha\beta\mu\nu}  \Gamma_{\mu\nu}^{\gamma\delta} u_{\gamma\delta} \equiv  - \tilde{\Lambda}^{\alpha\beta\gamma\delta} u_{\gamma\delta} \ ,
	\label{eq:Screeningrelation}
\end{equation}
where $\Lambda^{\alpha\beta\mu\nu}$ is the inverse of $\Lambda_{\alpha\beta\mu\nu}$.
Substituting in Eq.(\ref{StrainVar}) and integrating by parts we get
\begin{equation}
	\begin{split}
		\delta_d U &=  \int \dif^2 x \left( \sigma^{\alpha\beta} \partial_\alpha \delta d_\beta +   \Gamma^{\alpha\beta}_{\gamma\delta} Q^{\gamma\delta}\partial_\alpha \delta d_\beta\right) \\ &= \oint  \left(\sigma^{\alpha\beta} + \Gamma^{\alpha\beta}_{\gamma\delta} Q^{\gamma\delta}  \right) n_\alpha \delta d_\beta \dif \ell \\& -  \int \dif^2 x \partial_\alpha  \left(\sigma^{\alpha\beta} + \Gamma^{\alpha\beta}_{\gamma\delta} Q^{\gamma\delta}  \right)  \delta d_\beta 
	\end{split}
\end{equation} 
hence
\begin{equation}
	\partial_\alpha  \left(\sigma^{\alpha\beta} + \Gamma^{\alpha\beta}_{\gamma\delta} Q^{\gamma\delta}  \right)  = 0
	\label{eq:Equilibrium}
\end{equation}
Combining \eqref{eq:Screeningrelation} with \eqref{eq:Equilibrium} we find an equation of the form 
\begin{eqnarray}
	D^{\alpha\beta}_{\tau\mu}\partial_\alpha \sigma^{\tau\mu} = 0
\end{eqnarray}
with $D$ an invertible tensor depending on the coupling tensors $\Gamma,\Lambda,\A$. Upon inversion we get the standard equilibrium equation $\partial_{\alpha}\sigma^{\alpha\beta} = 0$.
The effective elastic properties are obtained by substituting the constitutive relation \eqref{eq:Screeningrelation} in the energy density \eqref{eq:Udecomp}.

\begin{equation}
	\begin{split}
		\Lag &=  \frac{1}{2} A^{\mu\nu\rho\sigma} u_{\mu\nu}u_{\rho\sigma} + 
		 \frac{1}{2} \Lambda_{\alpha\beta\gamma\delta} Q^{\alpha\beta}Q^{\gamma\delta}
		 + \Gamma^{\alpha\beta}_{\gamma\delta} u_{\alpha\beta}Q^{\gamma\delta} \\&=  \frac{1}{2} A^{\mu\nu\rho\sigma} u_{\mu\nu}u_{\rho\sigma} + 
		 \frac{1}{2} \Lambda_{\alpha\beta\gamma\delta} \left(- \tilde{\Lambda}^{\alpha\beta\mu\nu} u_{\mu\nu}\right) \left(- \tilde{\Lambda}^{\gamma\delta\rho\sigma} u_{\rho\sigma}\right)
		  \\&+ \Gamma^{\mu\nu}_{\gamma\delta} u_{\mu\nu}\left(- \tilde{\Lambda}^{\gamma\delta\rho\sigma} u_{\rho\sigma}\right) \equiv \frac{1}{2} \tilde{\A}^{\mu\nu\rho\sigma}u_{\mu\nu}u_{\rho\sigma}
	\end{split}
\end{equation} 
with 
\begin{eqnarray}
	\tilde{\A}^{\mu\nu\rho\sigma} = {\A}^{\mu\nu\rho\sigma} +  \Lambda_{\alpha\beta\gamma\delta}  \tilde{\Lambda}^{\alpha\beta\mu\nu}  \tilde{\Lambda}^{\gamma\delta\rho\sigma} - 2 \Gamma^{\mu\nu}_{\gamma\delta} \tilde{\Lambda}^{\gamma\delta\rho\sigma}
\end{eqnarray}
we see that the re-normalization of the quadrupole-quadrupole interactions results with a linear constitutive relation between inducing stress and induced quadrupoles which then renormalizes the elastic tensor \cite{20NWRZBC}. This is the analog of the situation in dielectrics where the dielectric constant is renormalized
by the induced dipoles. In the next subsection we provide examples of this situation.

\subsection{Numerical Examples} 
\label{exam1}

\begin{figure}
	\includegraphics[width=0.9\linewidth]{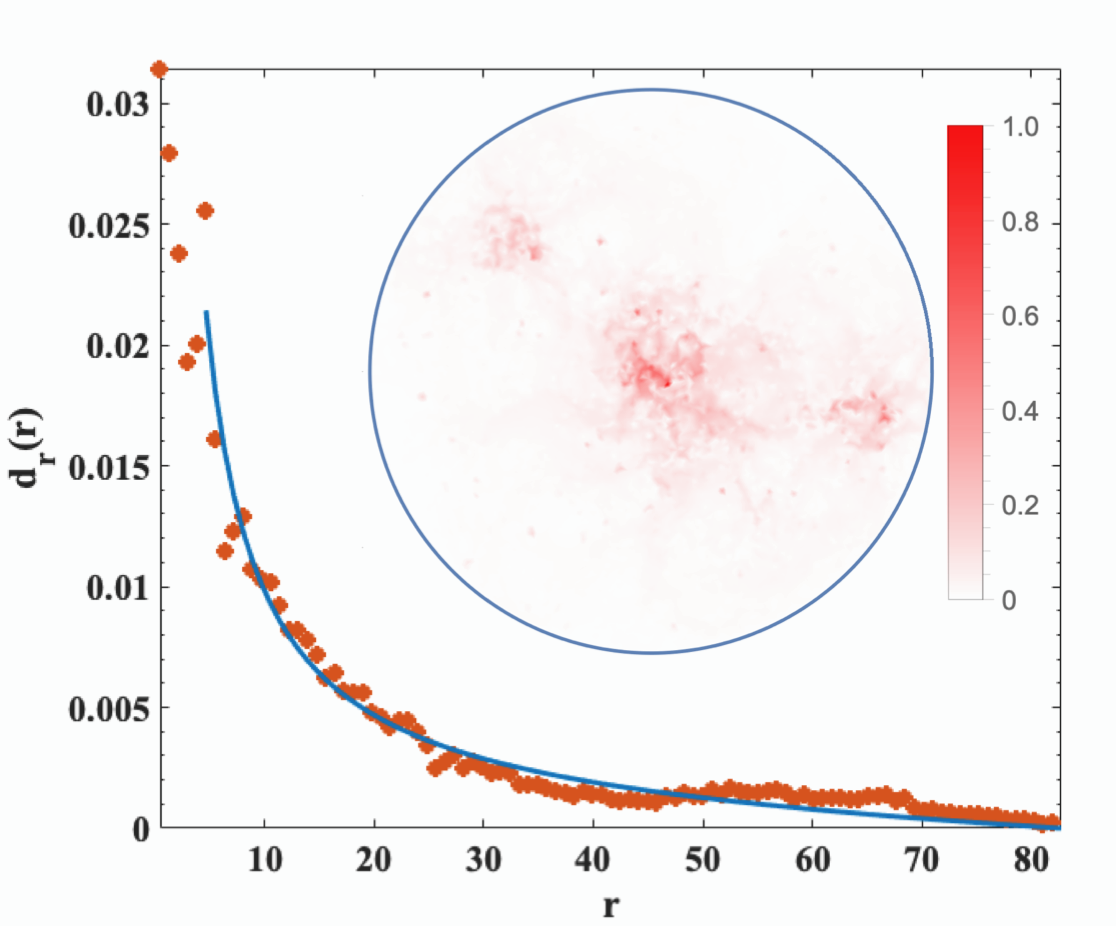}
	\caption{Displacement field for $P=40$. Inset: the displacement field caused by inflating one disk closest to the origin by 1\%. Here the displacement vectors are normalized and color coded as in Fig.~\ref{disP4.5}. Shown also is the fit of Eq.~(\ref{renelas}) to the angle-averaged displacement. Here the parameters used in the fit are $d_0=0.03028$; $\rin=3.268$; $\rout=82.74$.}
	\label{P40}
\end{figure}

\begin{figure}
		\includegraphics[width=1.0\linewidth]{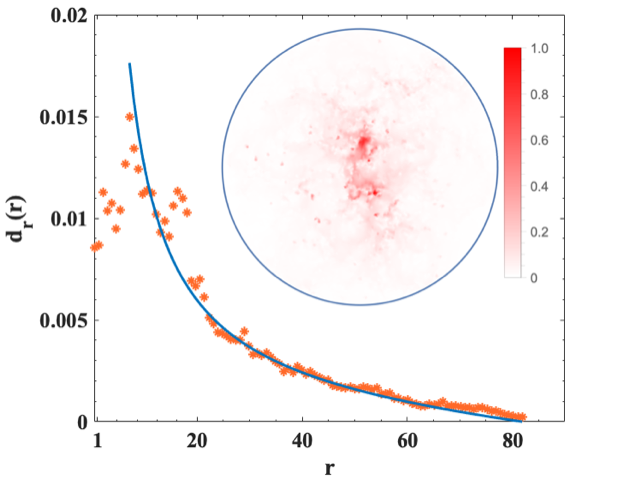}
		\caption{Displacement field for $P=60$. Inset: the displacement field caused by inflating one disk closest to the origin by 1\%. The displacement vectors are normalized and color coded as in Fig.~\ref{disP4.5} Shown also is the fit of Eq.~(\ref{renelas}) to the angle-averaged displacement. Here the parameters used in the fit are $d_0=0.03487$; $\rin=3.631$;$\rout=81.9$.}
		\label{P60}
	\end{figure}

To test our theory of the effect of low density quadrupoles we employ the same
system that gave rise to the results shown in Fig.~\ref{angav}, but at much higher
pressure, $P=40$ and $P=60$. We expect that at higher pressures the density of plastic events will be much lower, and indeed this is what we see, cf. the displacement maps in panel (a) of Figs.~\ref{P40} and \ref{P60}. Here the vectors of the displacement field are multiplied by a factor of 40 (compared to 10 for $P=4.5$. One can observe immediately that the displacement field is now concentrated near
the inflated disk and decays towards the outer boundary, as is expected in (renormalized) elasticity.

The angle-averaged displacement fields is computed from these maps, and are shown
together with the solution Eq.~(\ref{renelas}) in panels (b) of these figures.
We should stress that one does not expect agreement at low values of $r$ before the continuum limit that is used here becomes valid. We see that for $r\ge 10$ the continuum approximate fits extremely well the measured data. 

\section{Continuum Theory for higher densities of quadrupoles}
\label{robust}

\subsection{Constructing the theory}
The previous section indicated that we can continue our study following the common approach in constructing energy densities employing the natural interacting fields, by identifying the scalars that are consistent with the symmetries of the system. Equation \eqref{eq:Udecomp} is the most general quadratic energy that is isotropic and homogeneous, containing a density of quadrupoles interacting with the elastic fields of a solid. Motivated by the celebrated hexatic phase in 2d crystals \cite{80ZHN}, we now consider the case in which the pressure on the amorphous solid is sufficiently low (equivalent to high temperature in the 2d melting problem), such that quadrupoles assemble to dipoles. The quadrupoles tend to align in two dimensions, sometimes forming shear bands \cite{12DHP,13DHP}. In this case the amorphous solid is modeled as an elastic solid containing a density of quadrupoles and dipoles.

In the previous section we assumed that the density of quadrupoles is low, allowing us to neglect highr order terms in the multipole expansion, and in particular the gradient of the quadrupole density in the passage from Eq.~(\ref{defchi})to Eq.~(\ref{chicont}).
Presently we consider higher densities of quadrupoles, forcing us to add to the energy functional
gradients of the quadrupolar field.

To understand how dipoles are coupled to the elastic field we can look at the quadrupole-strain coupling and perform integration by parts to find
\begin{equation}
	\begin{split}
		&\int \dif^2 x \Gamma^{\alpha\beta}_{\gamma\delta} u_{\alpha\beta}Q^{\gamma\delta}  = \int \dif^2 x \Gamma^{\alpha\beta}_{\gamma\delta} \partial_\alpha d_{\beta}Q^{\gamma\delta} \\&= \oint \dif l\, \Gamma^{\alpha\beta}_{\gamma\delta}  d_{\beta}Q^{\gamma\delta} n_\alpha - \int \dif^2 x \Gamma^{\alpha\beta}_{\gamma\delta}  d_{\beta}\partial_\alpha Q^{\gamma\delta} \ .
	\end{split} 
\end{equation} 
This implies that the coupling of dipoles to elastic fields is via the displacement field. Since we found above that the effect of quadrupoles alone is limited to  renormalizing the elastic tensor, we can focus now on the consequences of the dipoles field. Denoting the dipole density $P^\alpha \equiv \partial_\beta Q^{\alpha\beta}$, the most general isotropic and homogeneous quadratic energy is 
\begin{equation}
	\begin{split}
		\Lag &=  \frac{1}{2} A^{\mu\nu\rho\sigma} u_{\mu\nu}u_{\rho\sigma} + 
		\frac{1}{2} \Lambda_{\alpha\beta} P^{\alpha}P^{\beta}
		+ \Gamma^{\alpha}_{\,\,\beta} d_{\alpha}P^{\beta} \ .
	\end{split}
\end{equation} 
At this point we should remind the reader that the notation using the dipole field $P^\alpha$ does not change the fact that the minimization of $\Lag$ should be done with respect that the fundamental field
$d_\alpha$ and $Q^{\alpha\beta}$. Minimization with respect to P would violate the conservation of dipoles. To make this point obvious we write the energy density in the explicit form  
\begin{equation}
	\begin{split}
		\Lag &=  \frac{1}{2} A^{\mu\nu\rho\sigma} u_{\mu\nu}u_{\rho\sigma} + 
		\frac{1}{2} \Lambda_{\alpha\beta} \partial_\mu Q^{\mu\alpha}  \partial_\nu Q^{\nu\beta}
		+ \Gamma_{\alpha}^{\,\,\beta} \partial_\mu Q^{\mu\alpha} d_{\beta} \ .
	\end{split}
\end{equation} 

Upon minimizing with respect to the fundamental fields $d$ and $Q$ we find 
\begin{equation}
	\begin{split}
		\delta_Q S = \delta_Q \int \Lag \dif^2 x=  \int \dif^2 x \left( \Lambda_{\alpha\beta} P^{\alpha}+ \Gamma^{\alpha}_{\beta} d_{\alpha}\right)\delta P^{\beta}\\
		\delta_d S = \delta_d \int \Lag \dif^2 x=  \int \dif^2 x \left( \sigma^{\alpha\beta}\delta u_{\alpha\beta} +   \Gamma^{\alpha}_{\beta} P^{\beta} \delta d_{\alpha}\right)  
	\end{split}
	\label{eq:StrainVar2}
\end{equation}

From the first equation we get a linear screening relation 
\begin{equation}
	P^{\alpha} = - \Lambda^{\alpha\beta}  \Gamma_{\beta}^{\gamma} d_{\gamma} \ .
	\label{eq:Screeningrelation2}
\end{equation}
In the second equation we use the relation between strain and displacement Eq.~(\ref{strainvar}).
Substituting Eq.~(\ref{strainvar}) and integrating by parts we get
\begin{equation}
	\begin{split}
		\delta_d U &=   \int \dif^2 x \left( \sigma^{\alpha\beta} \partial_\alpha \delta d_\beta    +   \Gamma_{\alpha}^{\beta} P^{\alpha} \delta d_{\beta}\right) \\ &= \oint  \sigma^{\alpha\beta} n_\alpha \delta d_\beta \dif \ell  \\&-  \int \dif^2 x   \left(\partial_\alpha \sigma^{\alpha\beta} + \Gamma_{\alpha}^{\beta} P^{\alpha} \right)  \delta d_\beta 
	\end{split}
\end{equation} 
hence
\begin{equation}
	\partial_\alpha \sigma^{\alpha\beta} =- \Gamma_{\alpha}^{\beta} P^{\alpha}
	\label{eq:Equilibrium2}
\end{equation}
Combining \eqref{eq:Screeningrelation2} with \eqref{eq:Equilibrium2} we find
\begin{equation}
	\partial_\alpha \sigma^{\alpha\beta} =- \Gamma_{\alpha}^{\beta} P^{\alpha} = \Gamma_{\alpha}^{\beta}  \Lambda^{\alpha\mu}  \Gamma_{\mu}^{\gamma} d_{\gamma}
	\label{eq:ScreeningEquilibrium2}
\end{equation}
We see that the displacement field acts as a screening source in the equilibrium equation. We  now rewrite this equation by substituting the stress in terms of strain, and the strain in terms of the displacement. In isotropic and homogeneous materials the coupling tensors have the following forms
\begin{eqnarray}
	\begin{split}
		\A^{\alpha\beta\gamma\delta} &= \lambda_1 g^{\alpha\beta} g^{\gamma\delta} + \lambda_2 \left(g^{\alpha\gamma} g^{\beta\delta}  + g^{\alpha\delta} g^{\beta\gamma} \right)\\
		\Gamma_{\alpha}^{\beta} &= \mu_1 g^\alpha_\beta\\
		\Lambda^{\alpha\beta} &= \mu_2 g^{\alpha\beta}
	\end{split}
\end{eqnarray}
with $\B g$ the euclidean metric tensor.
Direct substitution yields
\begin{equation}
	\lambda_2 \Delta \mathbf{d} + \left(\lambda_1+\lambda_2\right) \nabla \left(\nabla\cdot \mathbf{d}\right) = \mu_1 \mathbf{P} = -\frac{\mu_1^2}{\mu_2} \mathbf{d}
\end{equation}
or in a simpler form
\begin{equation}
	\Delta \mathbf{d} + \left(1+\frac{\lambda_1}{\lambda_2}\right) \nabla \left(\nabla\cdot \mathbf{d}\right) = -\frac{\mu_1^2}{\mu_2 \lambda_2 } \mathbf{d}
	\label{eq:DisplacementEquation}
\end{equation}
The screening effect is negligible when $\frac{\mu_1^2}{\mu_2 \lambda_2 } \ll 1$. At low pressures $\lambda_2 \to 0$ and the screening effect dominates.
Unlike the quadrupole screening, dipole screening leads to a qualitatively new behavior.

\subsection{Numerical Example}

\begin{figure}
	\centering
	\includegraphics[width=1.0\linewidth]{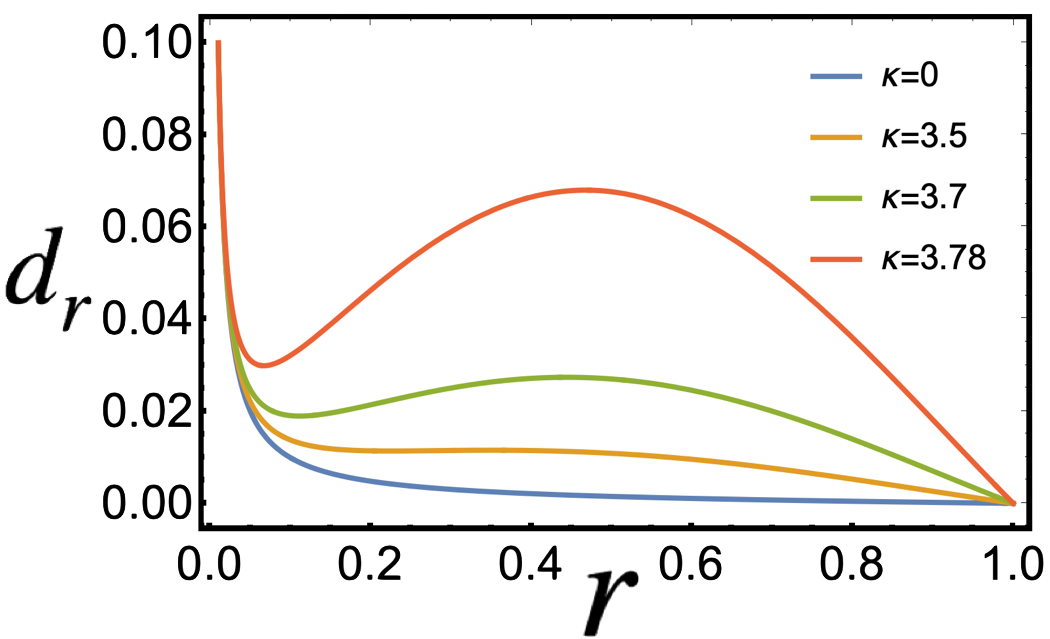}
	\caption{The solutions of Eq.~(\ref{amazing}) for different values of the parameter $\kappa$. }
	\label{solution}
	\end{figure}

Presently we discuss simulations that are identical in protocol and aim as those discussed in Subsect.~\ref{exam1}, but at much lower pressure, $P=4.5$. A disk which is closest to the origin was
inflated by 1\%.  Upon assuming polar symmetry Eq.~(\ref{eq:DisplacementEquation}) reduces to
\begin{equation}
	\Delta {\B d} = -\frac{\mu_1^2}{\mu_2 \left(\lambda_1 + 2\lambda_2\right)} {\B d} \equiv -\kappa^2 \, {\B d}\ .
	\label{eq:InclusionEquation}
\end{equation}
In polar coordinates 
\begin{equation}
	d_r'' +\frac{1}{r} d_r' +(\kappa^2 -\frac{1}{r^2})d_r=0 \ .
\end{equation}
This is the Bessel equation. A solution 
of this equation satisfying $d_r(r_{\rm in})=d_0$, $d_r(r_{\rm out})=0$ reads
\begin{equation}
	d_r(r)  = d_0 \frac{ Y_1(r \, \kappa ) J_1(r_\text{out} \kappa )-J_1(r \, \kappa ) Y_1(r_\text{out} \kappa )}{Y_1( r_\text{in} \kappa ) J_1(r_\text{out} \kappa )-J_1(r_\text{in} \kappa ) Y_1(r_\text{out} \kappa )} \ .
	\label{amazing}
\end{equation}
Here $J_1$ and $Y_1$ are the Bessel functions of the first and second kind respectively. The solution of this equation for different values of $\kappa$ are shown in Fig.~\ref{solution}.

\subsection{Comparison to the simulations}

The simulations of our circular box with configurations of binary disks serves
admirably to test the theory and the applicability of Eq.~(\ref{amazing}). Repeating the same
kind of simulations that led to Figs.~\ref{P40} and \ref{P60}, but at much lower
pressure $P=4.5$, we get typical displacement fields as seen in Fig.~\ref{disP4.5}. Fitting
Eq.~(\ref{amazing}) to the angle-averaged displacement field we get Fig.~\ref{angav}. Remembering that we should not expect a perfect fit of a continuum theory neither at $\rin$
nor at $\rout$, the quality of the fit seen is more than satisfactory. We thus propose
that the theory presented above appears quite relevant at least to model amorphous solids.
Of course, comparisons to experimental system is highly desirable and will be part
of our future research. 

\section{Plastic or elastic?}
\label{PorE}

\begin{figure}[h!]
	\centering
	\includegraphics[width=0.8\linewidth]{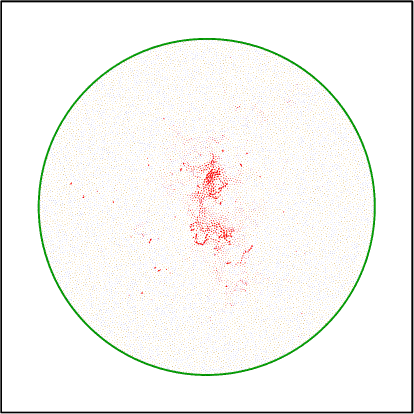}
	\includegraphics[width=0.8\linewidth]{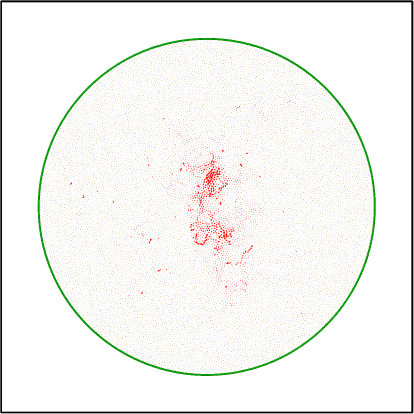}
	\caption{Comparison of typical displacement fields with respect to the unperturbed
	configuration, $P=60$. Upper panel: after inflation. Lower panel: after deflation.}
\label{test1}
\end{figure}

\begin{figure}[h!]
	\centering
	\includegraphics[width=0.8\linewidth]{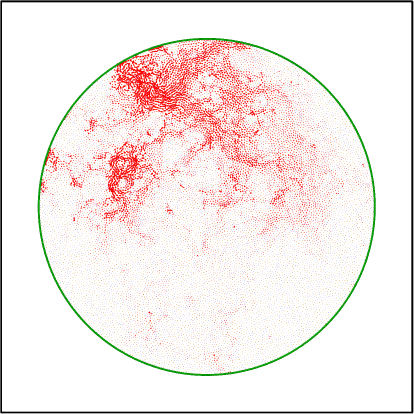}
	\includegraphics[width=0.8\linewidth]{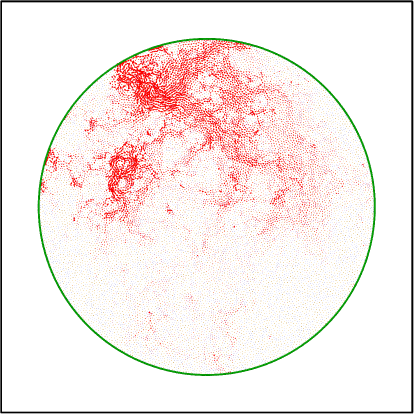}
	\caption{Comparison of typical displacement fields with respect to the unperturbed
		configuration, $P=4.5$. Upper panel: after inflation. Lower panel: after deflation.}
	\label{test2}
\end{figure}

Before summarizing the paper, we need to ascertain that the displacement fields shown in Figs.~\ref{disP4.5}, \ref{P40} and \ref{P60} are indeed resulting from plastic responses.
It is well known that upon applying strain to an amorphous solids the displacement fields are partly affine (following precisely the applied strain), and partly non-affine. The non-affine
responses can be reversible or irreversible. Only the latter can be referred to as plastic.
A straightforward way to test the reversibility of a measured displacement field is simply
to revert the applied perturbation. In the present cases, in which we always inflated
a central disk by 1\%, we can deflate it back, and measure the displacement field (after the
deflation) compared to the original configuration. We show two typical results of this test in 
Figs.~\ref{test1} and \ref{test2}. The first of these refers to the high pressure $P=60$. In the upper panel we display the displacement field after the inflation, and in the lower panel
after the deflation . We see that there is essentially no change.

The second of these tests is performed at $P=4.5$. The displacement fields after inflation
and after deflation are shown again in the upper and lower panels respectively, this
time in Fig. ~\ref{test2}. 

The conclusion is that upon deflation the displacement field is not annulled. In fact it remains invariant, showing unequivocally that we are dealing with non-affine plastic responses that are not reversible. This should be considered as a distinction from dielectrics, in which
dipoles annul when the electric field is switched off. 

\section{Summary and the road ahead}
\label{summary}

In summary, we have shown that the ever existing plastic responses in amorphous solids should be
carefully taken into account in assessing their implications for the mechanical properties of the host materials. The displacement field 
in response to a local source of stress can differ enormously from the expectation of classical
elasticity theory. Instead of decaying like $1/r^{D-1}$ the displacement field can increase, decrease, and oscillate as a function of $r$, and it is all dependent on the density of quadrupolar plastic
responses, their gradients (dipoles), and in extreme cases also monopoles. We have demonstrated
that measured displacement fields in model systems of amorphous configurations of binary disks in which one disk close to the origin is slightly inflated, agree excellently well with the theory
that was developed in this paper. 

It is worthwhile to reiterate that the kind of screening discussed here is analogous, but
in fact richer, than the electrostatic counterpart. While in electrostatics one is concerned
with charges and dipoles, here we have three possible screening agents, monopoles, dipoles
and quadrupoles. The case with small density of quadrupoles is analogous to dielectrics,
having only renormalization of the material parameters. The novel and interesting situation
of screening dipoles in elasticity, for which the displacement field changes qualitatively,
does not have an analog in electrostatics. Finally, monopoles may become relevant, but we did not deal with them explicitly since we expect that their appearance will be accompanied with
the melting (or un-jamming) of our amorphous solids.

Finally, we want to stress that the theory presented above is not limited to amorphous solids.
The way that the theory was constructed, using the scalar fields that are allowed by symmetry,
opens up the application of the theory to any situation in which elasticity is accompanied
by induced relaxation mechanisms. Such extensions will be provided in future publications. 

\acknowledgments
We acknowledge useful discussion with Prof. Konrad Samwer on the subject of this paper.
This work had been supported in part by the US-Israel Binational Science Foundation and the
Minerva Center for ``Aging, from physical materials to human tissues" at the Weizmann Institute. S.R. acknowledges the support of the Science and Engineering Research Board, DST, India under grant no. SRG/2020/001943 and the IIT Ropar under ISIRD grant. MM acknowledges support from the Israel Science Foundation (grant No. 1441/19).

\appendix

\section{System preparation and protocols}
\label{proto}

We investigate frictionless assemblies of circular disks that are at mechanical equilibrium,  prepared with a desired target pressure $P$ and confined in a circular two-dimensional box with a fixed outer wall. Open source codes (LAMMPS \cite{95Pli}) are used to perform the simulations. Every simulation begins with a dilute (area ratio $\phi=0.45$) configuration of $N$ =16000 bi-disperse disks placed randomly in a circular box with a radius, $\rout=80$ in SI units. Initially, half of the disks have a radius $R_1 =0.35$ and the other half a radius $R_2=0.49$, both in SI units. To reach a desired pressure the diameter of all the particles are inflated by a small factor($1.00004$) in each step, and the system is relaxed to mechanical equilibrium after every inflation step by solving Newton's second law of motion with damping. This process is carried out until the desired target pressure is reached.

The normal contact force is Hertzian, following  the Discrete Element Method of Ref.~\cite{79CS}. The tangential contact force is zero as the system is frictionless. Let us consider two particles $i$ and $j$, at positions $\B r_i$, $\B r_j$ with velocities $\B v_i$, $\B v_j$. Two particles interact if and only if there is an overlap i.e. if the relative normal compression $\Delta_{ij}^{(n)}=D_{ij}-r_{ij}>0$, where $r_{ij}=|\vec r_{ij}|$, $\B r_{ij}=\vec r_i-\vec r_j$, $D_{ij}=R_i+R_j$, and {$R_i$, $R_j$} the radii of disks $i$ and $j$. The normal unit vector is denoted as $\vec n_{ij}=\vec r_{ij}/r_{ij}$. Normal component of the relative velocity at contact is given as,
\begin{equation}
	\begin{split}
		{\B v}^{(n)}_{ij}&= ({\B v}_{ij} .\B n_{ij})\,\B n_{ij}  
	\end{split}
\end{equation}
The normal force exerted by grain $j$ on $i$ is
\begin{equation}
	\begin{split}
		\B F^{(n)}_{ij}&=k_n\Delta_{ij}^{(n)}\B n_{ij}-\frac{\gamma_n}{2} \B {v}^{(n)}_{ij}
	\end{split}
\end{equation}
where
\begin{equation}
	\begin{split}
		k_n &= k_n'\sqrt{ \Delta_{ij} R_{ij}} \ ,\\
		\gamma_{n} &= \gamma_{n}^{'}  \sqrt{ \Delta_{ij} R_{ij}}\ ,
	\end{split}
\end{equation}
with $R_{ij}^{-1}\equiv R_i^{-1}+R_j^{-1}$. $k_n^{'}$ is the normal spring stiffness. The parameter $\gamma_n^{'}$ is the viscoelastic damping constant.  
In the current simulations, the stiffness, $k_n=2\times 10^6$ N/m in SI units. The mass of each disk is $m=1$ in SI units. Each particle is inflated by a small factor followed by subsequent relaxation, annulling the total forces on each disk until they are smaller than $10^{-7}$ in SI units. This process is continued until a mechanically stable configuration is generated at a desired pressure $P$ in SI units.

After achieving a mechanically stable configurations at a target pressure, we choose the disk with larger diameter that is closest
to the center of the simulation box and inflate it by 1\%.  We then measure the displacement field that is induced by this inflation. This is the data shown
in throughout this paper.

\section{Microscopic derivation of Eqs.(\ref{eq:energydecomp})}.
\label{micro}

The detailed representation of the energy was presented in Ref.~\cite{13DHP}.
We show now that the present equations are equivalent. As an example consider the first term
in \eqref{eq:energydecomp}:
\begin{equation}
	\begin{split}
		U_\text{el} &=\int \dif^2 x \frac{1}{2} A^{\alpha\beta\gamma\delta} u_{\alpha\beta}(x)u_{\gamma\delta}(x) \\&=\int \dif^2 x \frac{1}{2} A^{\alpha\beta\gamma\delta} u^\text{el}_{\alpha\beta}(x) u^\text{el}_{\gamma\delta}(x)   \\&+ \int \dif^2 x  A^{\alpha\beta\gamma\delta}  u^\text{el}_{\alpha\beta}(x) u^\text{Q}_{\gamma\delta}(x)  \\& + \int \dif^2 x \frac{1}{2} A^{\alpha\beta\gamma\delta}  u^\text{Q}_{\alpha\beta}(x) u^\text{Q}_{\gamma\delta}(x)
	\end{split}
	\label{eq:Ueldecomp}
\end{equation} 
Here the first term corresponds to the self energy associated with the bare elastic fields. The second term 
vanishes identically. To see this we have to substitute the explicit expression for $G^u_{\gamma\delta\mu\nu}$ and perform integration by parts twice with respect to $x$ to obtain an integrand proportional to $\varepsilon^{\alpha\mu}\varepsilon^{\beta\nu} \partial_{\mu\nu} u^\text{el}_{\alpha\beta}$, which is the compatibility condition on the bare strain.

The last term describes the interactions between induced quadrupoles located at different points:
	\begin{eqnarray}
			&&\int \dif^2 x \frac{1}{2} A^{\alpha\beta\gamma\delta}  u^\text{Q}_{\alpha\beta}(x) u^\text{Q}_{\gamma\delta}(x)\nonumber\\&&  =  \int \dif^2 x \dif^2 x' \dif^2 x''  \frac{1}{2} A^{\alpha\beta\gamma\delta}  G_{\alpha\beta\mu\nu}^u(x-x')\nonumber\\&&\times G_{\gamma\delta\rho\sigma}^u(x-x'') Q^{\mu\nu}(x') Q^{\rho\sigma}(x'') \nonumber\\&&=  \int \dif^2 x' \dif^2 x''  \frac{1}{2}  Q^{\mu\nu}(x') Q^{\rho\sigma}(x'') \nonumber \\&&\times \int \dif^2 x A^{\alpha\beta\gamma\delta}  G_{\alpha\beta\mu\nu}^u(x-x') G_{\gamma\delta\rho\sigma}^u(x-x'') \nonumber \\&&\equiv   \int \dif^2 x' \dif^2 x''  \frac{1}{2}  Q^{\mu\nu}(x') Q^{\rho\sigma}(x'') \Lambda_{\mu\nu\rho\sigma} (x'-x'')
		\label{eq:uQuQ}
	\end{eqnarray} 

This expression is ill defined in the case $x'-x''\to 0$. Therefore renormalization techniques are required, where a cutoff length scale is introduced to regularize the integral, representing the quadrupoles core size. This result with an additional term describing quadrupoles self interactions, as in the second expression in \eqref{eq:energydecomp}. 

The third term in \eqref{eq:energydecomp} has two contributions. A quadrupole-quadrupole interaction term, correcting the coefficient of  \eqref{eq:uQuQ}, and a quadrupole-strain term
\begin{equation}
	\begin{split}
		U_\text{Q-el} &=  \int \dif^2 x \Gamma^{\alpha\beta}_{\gamma\delta} u_{\alpha\beta}Q^{\gamma\delta} =  \int \dif^2 x \Gamma^{\alpha\beta}_{\gamma\delta} u^\text{el}_{\alpha\beta}Q^{\gamma\delta} \\&+  \int \dif^2 x \dif^2 x' \Gamma^{\alpha\beta}_{\gamma\delta}G^ u_{\alpha\beta \mu\nu}(x-x')  Q^{\gamma\delta}(x) Q^{\mu\nu}(x')
	\end{split}
\end{equation} 
Wrapping it all together we find 
\begin{equation}
	\begin{split}
		U &=\int \dif^2 x \frac{1}{2} A^{\alpha\beta\gamma\delta} u^\text{el}_{\alpha\beta}(x) u^\text{el}_{\gamma\delta}(x)   \\& + \int \dif^2 x' \dif^2 x''  \frac{1}{2}  Q^{\mu\nu}(x') Q^{\rho\sigma}(x'') \Lambda_{\mu\nu\rho\sigma} (x'-x'') 
		\\&+\int \dif^2 x \frac{1}{2} \Lambda_{\alpha\beta\gamma\delta} Q^{\alpha\beta}Q^{\gamma\delta} +\int \dif^2 x \Gamma^{\alpha\beta}_{\gamma\delta} u^\text{el}_{\alpha\beta}Q^{\gamma\delta}
	\end{split}
\end{equation} 
This is exactly the form of the energy functional in \cite{13DHP}, and therefore we focus on the functional form  in \eqref{eq:Udecomp},\eqref{eq:energydecomp}.

\bibliography{ALL.anomalous}

\begin{thebibliography}{23}%
\makeatletter
\providecommand \@ifxundefined [1]{%
 \@ifx{#1\undefined}
}%
\providecommand \@ifnum [1]{%
 \ifnum #1\expandafter \@firstoftwo
 \else \expandafter \@secondoftwo
 \fi
}%
\providecommand \@ifx [1]{%
 \ifx #1\expandafter \@firstoftwo
 \else \expandafter \@secondoftwo
 \fi
}%
\providecommand \natexlab [1]{#1}%
\providecommand \enquote  [1]{``#1''}%
\providecommand \bibnamefont  [1]{#1}%
\providecommand \bibfnamefont [1]{#1}%
\providecommand \citenamefont [1]{#1}%
\providecommand \href@noop [0]{\@secondoftwo}%
\providecommand \href [0]{\begingroup \@sanitize@url \@href}%
\providecommand \@href[1]{\@@startlink{#1}\@@href}%
\providecommand \@@href[1]{\endgroup#1\@@endlink}%
\providecommand \@sanitize@url [0]{\catcode `\\12\catcode `\$12\catcode
  `\&12\catcode `\#12\catcode `\^12\catcode `\_12\catcode `\%12\relax}%
\providecommand \@@startlink[1]{}%
\providecommand \@@endlink[0]{}%
\providecommand \url  [0]{\begingroup\@sanitize@url \@url }%
\providecommand \@url [1]{\endgroup\@href {#1}{\urlprefix }}%
\providecommand \urlprefix  [0]{URL }%
\providecommand \Eprint [0]{\href }%
\providecommand \doibase [0]{https://doi.org/}%
\providecommand \selectlanguage [0]{\@gobble}%
\providecommand \bibinfo  [0]{\@secondoftwo}%
\providecommand \bibfield  [0]{\@secondoftwo}%
\providecommand \translation [1]{[#1]}%
\providecommand \BibitemOpen [0]{}%
\providecommand \bibitemStop [0]{}%
\providecommand \bibitemNoStop [0]{.\EOS\space}%
\providecommand \EOS [0]{\spacefactor3000\relax}%
\providecommand \BibitemShut  [1]{\csname bibitem#1\endcsname}%
\let\auto@bib@innerbib\@empty
\bibitem [{\citenamefont {LIFSHITZ}\ \emph {et~al.}(1986)\citenamefont
  {LIFSHITZ}, \citenamefont {KOSEVICH},\ and\ \citenamefont
  {PITAEVSKII}}]{86LKP}%
  \BibitemOpen
  \bibfield  {author} {\bibinfo {author} {\bibfnamefont {E.}~\bibnamefont
  {LIFSHITZ}}, \bibinfo {author} {\bibfnamefont {A.}~\bibnamefont {KOSEVICH}},\
  and\ \bibinfo {author} {\bibfnamefont {L.}~\bibnamefont {PITAEVSKII}},\
  }\bibfield  {title} {\bibinfo {title} {Chapter i - fundamental equations},\
  }in\ \href
  {https://doi.org/https://doi.org/10.1016/B978-0-08-057069-3.50008-5} {\emph
  {\bibinfo {booktitle} {Theory of Elasticity (Third Edition)}}},\ \bibinfo
  {editor} {edited by\ \bibinfo {editor} {\bibfnamefont {E.}~\bibnamefont
  {LIFSHITZ}}, \bibinfo {editor} {\bibfnamefont {A.}~\bibnamefont {KOSEVICH}},\
  and\ \bibinfo {editor} {\bibfnamefont {L.}~\bibnamefont {PITAEVSKII}}}\
  (\bibinfo  {publisher} {Butterworth-Heinemann},\ \bibinfo {address}
  {Oxford},\ \bibinfo {year} {1986})\ \bibinfo {edition} {third edition}\ ed.,\
  pp.\ \bibinfo {pages} {1--37}\BibitemShut {NoStop}%
\bibitem [{\citenamefont {Malandro}\ and\ \citenamefont {Lacks}(1999)}]{99ML}%
  \BibitemOpen
  \bibfield  {author} {\bibinfo {author} {\bibfnamefont {D.~L.}\ \bibnamefont
  {Malandro}}\ and\ \bibinfo {author} {\bibfnamefont {D.~J.}\ \bibnamefont
  {Lacks}},\ }\bibfield  {title} {\bibinfo {title} {{Relationships of
  shear-induced changes in the potential energy landscape to the mechanical
  properties of ductile glasses}},\ }\href {https://doi.org/10.1063/1.478340}
  {\bibfield  {journal} {\bibinfo  {journal} {J. Chem. Phys}\ ,\ \bibinfo
  {pages} {4593}} (\bibinfo {year} {1999})}\BibitemShut {NoStop}%
\bibitem [{\citenamefont {Maloney}\ and\ \citenamefont
  {Lema\^{\i}tre}(2006)}]{06ML}%
  \BibitemOpen
  \bibfield  {author} {\bibinfo {author} {\bibfnamefont {C.~E.}\ \bibnamefont
  {Maloney}}\ and\ \bibinfo {author} {\bibfnamefont {A.}~\bibnamefont
  {Lema\^{\i}tre}},\ }\bibfield  {title} {\bibinfo {title} {Amorphous systems
  in athermal, quasistatic shear},\ }\href
  {https://doi.org/10.1103/PhysRevE.74.016118} {\bibfield  {journal} {\bibinfo
  {journal} {Phys. Rev. E}\ }\textbf {\bibinfo {volume} {74}},\ \bibinfo
  {pages} {016118} (\bibinfo {year} {2006})}\BibitemShut {NoStop}%
\bibitem [{\citenamefont {Karmakar}\ \emph {et~al.}(2010)\citenamefont
  {Karmakar}, \citenamefont {Lerner},\ and\ \citenamefont
  {Procaccia}}]{10KLPb}%
  \BibitemOpen
  \bibfield  {author} {\bibinfo {author} {\bibfnamefont {S.}~\bibnamefont
  {Karmakar}}, \bibinfo {author} {\bibfnamefont {E.}~\bibnamefont {Lerner}},\
  and\ \bibinfo {author} {\bibfnamefont {I.}~\bibnamefont {Procaccia}},\
  }\bibfield  {title} {\bibinfo {title} {Statistical physics of the yielding
  transition in amorphous solids},\ }\href
  {https://doi.org/10.1103/PhysRevE.82.055103} {\bibfield  {journal} {\bibinfo
  {journal} {Phys. Rev. E}\ }\textbf {\bibinfo {volume} {82}},\ \bibinfo
  {pages} {055103} (\bibinfo {year} {2010})}\BibitemShut {NoStop}%
\bibitem [{\citenamefont {Hentschel}\ \emph {et~al.}(2011)\citenamefont
  {Hentschel}, \citenamefont {Karmakar}, \citenamefont {Lerner},\ and\
  \citenamefont {Procaccia}}]{11HKLP}%
  \BibitemOpen
  \bibfield  {author} {\bibinfo {author} {\bibfnamefont {H.~G.~E.}\
  \bibnamefont {Hentschel}}, \bibinfo {author} {\bibfnamefont {S.}~\bibnamefont
  {Karmakar}}, \bibinfo {author} {\bibfnamefont {E.}~\bibnamefont {Lerner}},\
  and\ \bibinfo {author} {\bibfnamefont {I.}~\bibnamefont {Procaccia}},\
  }\bibfield  {title} {\bibinfo {title} {{Do athermal amorphous solids
  exist?}},\ }\href {https://doi.org/10.1103/PhysRevE.83.061101} {\bibfield
  {journal} {\bibinfo  {journal} {Phys. Rev.E}\ }\textbf {\bibinfo {volume}
  {83}},\ \bibinfo {pages} {061101} (\bibinfo {year} {2011})}\BibitemShut
  {NoStop}%
\bibitem [{\citenamefont {Procaccia}\ \emph {et~al.}(2016)\citenamefont
  {Procaccia}, \citenamefont {Rainone}, \citenamefont {Shor},\ and\
  \citenamefont {Singh}}]{16PRS}%
  \BibitemOpen
  \bibfield  {author} {\bibinfo {author} {\bibfnamefont {I.}~\bibnamefont
  {Procaccia}}, \bibinfo {author} {\bibfnamefont {C.}~\bibnamefont {Rainone}},
  \bibinfo {author} {\bibfnamefont {C.~A. B.~Z.}\ \bibnamefont {Shor}},\ and\
  \bibinfo {author} {\bibfnamefont {M.}~\bibnamefont {Singh}},\ }\bibfield
  {title} {\bibinfo {title} {Breakdown of nonlinear elasticity in amorphous
  solids at finite temperatures},\ }\href
  {https://doi.org/10.1103/PhysRevE.93.063003} {\bibfield  {journal} {\bibinfo
  {journal} {Phys. Rev. E}\ }\textbf {\bibinfo {volume} {93}},\ \bibinfo
  {pages} {063003} (\bibinfo {year} {2016})}\BibitemShut {NoStop}%
\bibitem [{\citenamefont {Dubey}\ \emph {et~al.}(2016)\citenamefont {Dubey},
  \citenamefont {Procaccia}, \citenamefont {Shor},\ and\ \citenamefont
  {Singh}}]{16DPSS}%
  \BibitemOpen
  \bibfield  {author} {\bibinfo {author} {\bibfnamefont {A.~K.}\ \bibnamefont
  {Dubey}}, \bibinfo {author} {\bibfnamefont {I.}~\bibnamefont {Procaccia}},
  \bibinfo {author} {\bibfnamefont {C.~A.}\ \bibnamefont {Shor}},\ and\
  \bibinfo {author} {\bibfnamefont {M.}~\bibnamefont {Singh}},\ }\bibfield
  {title} {\bibinfo {title} {{Elasticity in Amorphous Solids: Nonlinear or
  Piecewise Linear?}},\ }\href {https://doi.org/10.1103/PhysRevLett.116.085502}
  {\bibfield  {journal} {\bibinfo  {journal} {Phys. Rev. Lett.}\ }\textbf
  {\bibinfo {volume} {116}},\ \bibinfo {pages} {085502} (\bibinfo {year}
  {2016})}\BibitemShut {NoStop}%
\bibitem [{\citenamefont {Dailidonis}\ \emph {et~al.}(2017)\citenamefont
  {Dailidonis}, \citenamefont {Ilyin}, \citenamefont {Procaccia},\ and\
  \citenamefont {Shor}}]{17DIPS}%
  \BibitemOpen
  \bibfield  {author} {\bibinfo {author} {\bibfnamefont {V.}~\bibnamefont
  {Dailidonis}}, \bibinfo {author} {\bibfnamefont {V.}~\bibnamefont {Ilyin}},
  \bibinfo {author} {\bibfnamefont {I.}~\bibnamefont {Procaccia}},\ and\
  \bibinfo {author} {\bibfnamefont {C.~A. B.~Z.}\ \bibnamefont {Shor}},\
  }\bibfield  {title} {\bibinfo {title} {Breakdown of nonlinear elasticity in
  stress-controlled thermal amorphous solids},\ }\href
  {https://doi.org/10.1103/PhysRevE.95.031001} {\bibfield  {journal} {\bibinfo
  {journal} {Phys. Rev. E}\ }\textbf {\bibinfo {volume} {95}},\ \bibinfo
  {pages} {031001} (\bibinfo {year} {2017})}\BibitemShut {NoStop}%
\bibitem [{\citenamefont {Dasgupta}\ \emph {et~al.}(2012)\citenamefont
  {Dasgupta}, \citenamefont {Hentschel},\ and\ \citenamefont
  {Procaccia}}]{12DHP}%
  \BibitemOpen
  \bibfield  {author} {\bibinfo {author} {\bibfnamefont {R.}~\bibnamefont
  {Dasgupta}}, \bibinfo {author} {\bibfnamefont {H.~G.~E.}\ \bibnamefont
  {Hentschel}},\ and\ \bibinfo {author} {\bibfnamefont {I.}~\bibnamefont
  {Procaccia}},\ }\bibfield  {title} {\bibinfo {title} {{Microscopic mechanism
  of shear bands in amorphous solids}},\ }\href
  {https://link.aps.org/doi/10.1103/PhysRevLett.109.255502} {\bibfield
  {journal} {\bibinfo  {journal} {Phys. Rev. Lett.}\ }\textbf {\bibinfo
  {volume} {109}},\ \bibinfo {pages} {255502} (\bibinfo {year}
  {2012})}\BibitemShut {NoStop}%
\bibitem [{\citenamefont {Dasgupta}\ \emph {et~al.}(2013)\citenamefont
  {Dasgupta}, \citenamefont {Hentschel},\ and\ \citenamefont
  {Procaccia}}]{13DHP}%
  \BibitemOpen
  \bibfield  {author} {\bibinfo {author} {\bibfnamefont {R.}~\bibnamefont
  {Dasgupta}}, \bibinfo {author} {\bibfnamefont {H.~G.~E.}\ \bibnamefont
  {Hentschel}},\ and\ \bibinfo {author} {\bibfnamefont {I.}~\bibnamefont
  {Procaccia}},\ }\bibfield  {title} {\bibinfo {title} {{Yield strain in shear
  banding amorphous solids}},\ }\href
  {https://link.aps.org/doi/10.1103/PhysRevE.87.022810} {\bibfield  {journal}
  {\bibinfo  {journal} {Phys. Rev.E}\ }\textbf {\bibinfo {volume} {87}},\
  \bibinfo {pages} {022810} (\bibinfo {year} {2013})}\BibitemShut {NoStop}%
\bibitem [{\citenamefont {Landau}\ and\ \citenamefont {Lifshitz}(1984)}]{84LL}%
  \BibitemOpen
  \bibfield  {author} {\bibinfo {author} {\bibfnamefont {L.}~\bibnamefont
  {Landau}}\ and\ \bibinfo {author} {\bibfnamefont {E.}~\bibnamefont
  {Lifshitz}},\ }\bibfield  {title} {\bibinfo {title} {Chapter
  i--electrostatics of conductors},\ }\href@noop {} {\bibfield  {journal}
  {\bibinfo  {journal} {Electrodynamics of Continuous Media (Second Edition
  Revised and Enlarged)}\ ,\ \bibinfo {pages} {1}} (\bibinfo {year}
  {1984})}\BibitemShut {NoStop}%
\bibitem [{\citenamefont {Lema\^{\i}tre}\ \emph {et~al.}(2021)\citenamefont
  {Lema\^{\i}tre}, \citenamefont {Mondal}, \citenamefont {Procaccia},\ and\
  \citenamefont {Roy}}]{21LMPR}%
  \BibitemOpen
  \bibfield  {author} {\bibinfo {author} {\bibfnamefont {A.}~\bibnamefont
  {Lema\^{\i}tre}}, \bibinfo {author} {\bibfnamefont {C.}~\bibnamefont
  {Mondal}}, \bibinfo {author} {\bibfnamefont {I.}~\bibnamefont {Procaccia}},\
  and\ \bibinfo {author} {\bibfnamefont {S.}~\bibnamefont {Roy}},\ }\bibfield
  {title} {\bibinfo {title} {Stress correlations in frictional granular
  media},\ }\href {https://doi.org/10.1103/PhysRevB.103.054110} {\bibfield
  {journal} {\bibinfo  {journal} {Phys. Rev. B}\ }\textbf {\bibinfo {volume}
  {103}},\ \bibinfo {pages} {054110} (\bibinfo {year} {2021})}\BibitemShut
  {NoStop}%
\bibitem [{\citenamefont {Nampoothiri}\ \emph {et~al.}(2020)\citenamefont
  {Nampoothiri}, \citenamefont {Wang}, \citenamefont {Ramola}, \citenamefont
  {Zhang}, \citenamefont {Bhattacharjee},\ and\ \citenamefont
  {Chakraborty}}]{20NWRZBC}%
  \BibitemOpen
  \bibfield  {author} {\bibinfo {author} {\bibfnamefont {J.~N.}\ \bibnamefont
  {Nampoothiri}}, \bibinfo {author} {\bibfnamefont {Y.}~\bibnamefont {Wang}},
  \bibinfo {author} {\bibfnamefont {K.}~\bibnamefont {Ramola}}, \bibinfo
  {author} {\bibfnamefont {J.}~\bibnamefont {Zhang}}, \bibinfo {author}
  {\bibfnamefont {S.}~\bibnamefont {Bhattacharjee}},\ and\ \bibinfo {author}
  {\bibfnamefont {B.}~\bibnamefont {Chakraborty}},\ }\bibfield  {title}
  {\bibinfo {title} {Emergent elasticity in amorphous solids},\ }\href
  {https://doi.org/10.1103/PhysRevLett.125.118002} {\bibfield  {journal}
  {\bibinfo  {journal} {Phys. Rev. Lett.}\ }\textbf {\bibinfo {volume} {125}},\
  \bibinfo {pages} {118002} (\bibinfo {year} {2020})}\BibitemShut {NoStop}%
\bibitem [{\citenamefont {Zippelius}\ \emph {et~al.}(1980)\citenamefont
  {Zippelius}, \citenamefont {Halperin},\ and\ \citenamefont {Nelson}}]{80ZHN}%
  \BibitemOpen
  \bibfield  {author} {\bibinfo {author} {\bibfnamefont {A.}~\bibnamefont
  {Zippelius}}, \bibinfo {author} {\bibfnamefont {B.~I.}\ \bibnamefont
  {Halperin}},\ and\ \bibinfo {author} {\bibfnamefont {D.~R.}\ \bibnamefont
  {Nelson}},\ }\bibfield  {title} {\bibinfo {title} {Dynamics of
  two-dimensional melting},\ }\href {https://doi.org/10.1103/PhysRevB.22.2514}
  {\bibfield  {journal} {\bibinfo  {journal} {Phys. Rev. B}\ }\textbf {\bibinfo
  {volume} {22}},\ \bibinfo {pages} {2514} (\bibinfo {year}
  {1980})}\BibitemShut {NoStop}%
\bibitem [{\citenamefont {Gurtin}(1973)}]{73Gur}%
  \BibitemOpen
  \bibfield  {author} {\bibinfo {author} {\bibfnamefont {M.~E.}\ \bibnamefont
  {Gurtin}},\ }\bibfield  {title} {\bibinfo {title} {The linear theory of
  elasticity},\ }in\ \href@noop {} {\emph {\bibinfo {booktitle} {Linear
  theories of elasticity and thermoelasticity}}}\ (\bibinfo  {publisher}
  {Springer},\ \bibinfo {year} {1973})\ pp.\ \bibinfo {pages}
  {1--295}\BibitemShut {NoStop}%
\bibitem [{\citenamefont {Muskhelishvili}()}]{53Mus}%
  \BibitemOpen
  \bibfield  {author} {\bibinfo {author} {\bibfnamefont {N.}~\bibnamefont
  {Muskhelishvili}},\ }\href@noop {} {\emph {\bibinfo {title} {Some basic
  problems of the mathematical theory of elasticity}}}\BibitemShut {NoStop}%
\bibitem [{\citenamefont {Eshelby}(1957)}]{54Esh}%
  \BibitemOpen
  \bibfield  {author} {\bibinfo {author} {\bibfnamefont {J.~D.}\ \bibnamefont
  {Eshelby}},\ }\bibfield  {title} {\bibinfo {title} {The determination of the
  elastic field of an ellipsoidal inclusion, and related problems},\ }\href
  {https://doi.org/10.1098/rspa.1957.0133} {\bibfield  {journal} {\bibinfo
  {journal} {Proceedings of the Royal Society of London A: Mathematical,
  Physical and Engineering Sciences}\ }\textbf {\bibinfo {volume} {241}},\
  \bibinfo {pages} {376} (\bibinfo {year} {1957})}\BibitemShut {NoStop}%
\bibitem [{\citenamefont {Moshe}\ \emph {et~al.}(2015)\citenamefont {Moshe},
  \citenamefont {Sharon},\ and\ \citenamefont {Kupferman}}]{15MSK}%
  \BibitemOpen
  \bibfield  {author} {\bibinfo {author} {\bibfnamefont {M.}~\bibnamefont
  {Moshe}}, \bibinfo {author} {\bibfnamefont {E.}~\bibnamefont {Sharon}},\ and\
  \bibinfo {author} {\bibfnamefont {R.}~\bibnamefont {Kupferman}},\ }\bibfield
  {title} {\bibinfo {title} {Elastic interactions between two-dimensional
  geometric defects},\ }\href@noop {} {\bibfield  {journal} {\bibinfo
  {journal} {Physical Review E}\ }\textbf {\bibinfo {volume} {92}},\ \bibinfo
  {pages} {062403} (\bibinfo {year} {2015})}\BibitemShut {NoStop}%
\bibitem [{\citenamefont {Sarkar}\ \emph
  {et~al.}(2021{\natexlab{a}})\citenamefont {Sarkar}, \citenamefont
  {{\v{C}}ebron}, \citenamefont {Brojan},\ and\ \citenamefont
  {Ko{\v{s}}mrlj}}]{21SCBK}%
  \BibitemOpen
  \bibfield  {author} {\bibinfo {author} {\bibfnamefont {S.}~\bibnamefont
  {Sarkar}}, \bibinfo {author} {\bibfnamefont {M.}~\bibnamefont
  {{\v{C}}ebron}}, \bibinfo {author} {\bibfnamefont {M.}~\bibnamefont
  {Brojan}},\ and\ \bibinfo {author} {\bibfnamefont {A.}~\bibnamefont
  {Ko{\v{s}}mrlj}},\ }\bibfield  {title} {\bibinfo {title} {Elastic multipole
  method for describing deformation of infinite two-dimensional solids with
  circular inclusions},\ }\href@noop {} {\bibfield  {journal} {\bibinfo
  {journal} {Physical Review E}\ }\textbf {\bibinfo {volume} {103}},\ \bibinfo
  {pages} {053003} (\bibinfo {year} {2021}{\natexlab{a}})}\BibitemShut
  {NoStop}%
\bibitem [{\citenamefont {Sarkar}\ \emph
  {et~al.}(2021{\natexlab{b}})\citenamefont {Sarkar}, \citenamefont
  {{\v{C}}ebron}, \citenamefont {Brojan},\ and\ \citenamefont
  {Ko{\v{s}}mrlj}}]{21SCBKa}%
  \BibitemOpen
  \bibfield  {author} {\bibinfo {author} {\bibfnamefont {S.}~\bibnamefont
  {Sarkar}}, \bibinfo {author} {\bibfnamefont {M.}~\bibnamefont
  {{\v{C}}ebron}}, \bibinfo {author} {\bibfnamefont {M.}~\bibnamefont
  {Brojan}},\ and\ \bibinfo {author} {\bibfnamefont {A.}~\bibnamefont
  {Ko{\v{s}}mrlj}},\ }\bibfield  {title} {\bibinfo {title} {Method of image
  charges for describing deformation of bounded two-dimensional solids with
  circular inclusions},\ }\href@noop {} {\bibfield  {journal} {\bibinfo
  {journal} {Physical Review E}\ }\textbf {\bibinfo {volume} {103}},\ \bibinfo
  {pages} {053004} (\bibinfo {year} {2021}{\natexlab{b}})}\BibitemShut
  {NoStop}%
\bibitem [{\citenamefont {Bar-Sinai}\ \emph {et~al.}(2020)\citenamefont
  {Bar-Sinai}, \citenamefont {Librandi}, \citenamefont {Bertoldi},\ and\
  \citenamefont {Moshe}}]{20BLBM}%
  \BibitemOpen
  \bibfield  {author} {\bibinfo {author} {\bibfnamefont {Y.}~\bibnamefont
  {Bar-Sinai}}, \bibinfo {author} {\bibfnamefont {G.}~\bibnamefont {Librandi}},
  \bibinfo {author} {\bibfnamefont {K.}~\bibnamefont {Bertoldi}},\ and\
  \bibinfo {author} {\bibfnamefont {M.}~\bibnamefont {Moshe}},\ }\bibfield
  {title} {\bibinfo {title} {Geometric charges and nonlinear elasticity of
  two-dimensional elastic metamaterials},\ }\href@noop {} {\bibfield  {journal}
  {\bibinfo  {journal} {Proceedings of the National Academy of Sciences}\
  }\textbf {\bibinfo {volume} {117}},\ \bibinfo {pages} {10195} (\bibinfo
  {year} {2020})}\BibitemShut {NoStop}%
\bibitem [{\citenamefont {Plimpton}(1995)}]{95Pli}%
  \BibitemOpen
  \bibfield  {author} {\bibinfo {author} {\bibfnamefont {S.}~\bibnamefont
  {Plimpton}},\ }\bibfield  {title} {\bibinfo {title} {{Fast Parallel
  Algorithms for Short-Range Molecular Dynamics}},\ }\href
  {https://doi.org/10.1006/jcph.1995.1039} {\bibfield  {journal} {\bibinfo
  {journal} {Journal of Computational Physics}\ }\textbf {\bibinfo {volume}
  {117}},\ \bibinfo {pages} {1} (\bibinfo {year} {1995})}\BibitemShut {NoStop}%
\bibitem [{\citenamefont {Cundall}\ and\ \citenamefont {Strack}(1979)}]{79CS}%
  \BibitemOpen
  \bibfield  {author} {\bibinfo {author} {\bibfnamefont {P.~A.}\ \bibnamefont
  {Cundall}}\ and\ \bibinfo {author} {\bibfnamefont {O.~D.~L.}\ \bibnamefont
  {Strack}},\ }\bibfield  {title} {\bibinfo {title} {A discrete numerical model
  for granular assemblies},\ }\href {https://doi.org/10.1680/geot.1979.29.1.47}
  {\bibfield  {journal} {\bibinfo  {journal} {Géotechnique}\ }\textbf
  {\bibinfo {volume} {29}},\ \bibinfo {pages} {47} (\bibinfo {year}
  {1979})}\BibitemShut {NoStop}%
\end{thebibliography}%

\end{document}